\documentclass[5p,twocolumn, times]{elsarticle}
\usepackage[utf8]{inputenc}
\usepackage{braket}
\usepackage{graphicx}
\usepackage{natbib}
\biboptions{sort&compress}
\usepackage{xcolor}
\usepackage[normalem]{ulem}
\usepackage[colorlinks=true,linkcolor=black, citecolor=blue, urlcolor=blue]{hyperref}

\begin{document}
\begin{frontmatter}
\title{Particle Track Classification Using Quantum Associative Memory}

\author[1]{Gregory Quiroz}
\author[1]{Lauren Ice}
\author[2]{Andrea Delgado}
\author[3,4]{Travis S.~Humble}

\address[1]{The Johns Hopkins University Applied Physics Laboratory, \\
11100 Johns Hopkins Road, Laurel, MD, 20723, USA}
\address[2]{Physics Division, Oak Ridge National Laboratory, Oak Ridge, TN, 37831, USA}
\address[3]{Quantum Computing Institute, Oak Ridge National Laboratory, Oak Ridge, TN, 37831, USA}
\address[4]{Computational Sciences and Engineering, Oak Ridge National Laboratory, Oak Ridge, TN, 37831, USA}

	
\begin{abstract}
Pattern recognition algorithms are commonly employed to simplify the challenging and necessary step of track reconstruction in sub-atomic physics experiments.
Aiding in the discrimination of relevant interactions, pattern recognition seeks to accelerate track reconstruction by isolating signals
of interest.
In high collision rate experiments, such algorithms can be particularly crucial for determining whether to retain or discard information from a given interaction even before the data is transferred to tape. As data rates, detector resolution, noise, and inefficiencies increase, pattern recognition becomes more computationally challenging, motivating the development of higher efficiency algorithms and techniques. Quantum associative memory is an approach that seeks to exploits quantum mechanical phenomena to gain advantage in learning capacity, or the number of patterns that can be stored and accurately recalled. Here, we study quantum associative memory based on quantum annealing and apply it to the particle track classification. We focus on discrimination models based on Ising formulations of quantum associative memory model (QAMM) recall and quantum content-addressable memory (QCAM) recall. We characterize classification performance of these approaches as a function detector resolution, pattern library size, and detector inefficiencies, using the D-Wave 2000Q processor as a testbed. Discrimination criteria is set using both solution-state energy and classification labels embedded in solution states.
We find that energy-based QAMM classification performs well in regimes of small pattern density and low detector inefficiency. In contrast, state-based QCAM achieves reasonably high accuracy recall for large pattern density and the greatest recall accuracy robustness to a variety of detector noise sources. 
\end{abstract}
\end{frontmatter}

\section{Introduction}
Sub-atomic physics experiments are designed to extract information about the interactions and properties of sub-atomic particles. The process often involves reconstructing the trajectory (known as a track) of a charged particle through a detector from which the point of origin, charge, and momentum can be determined. Tracks are reconstructed from the spatial coordinates of signals in the detector, commonly referred to as ``hits'' produced when the charged particle interacts with detector materials. In experiments with simple detector geometries, low noise, high efficiencies, and low track multiplicity, particle trajectories can be quickly determined from the observed detector hits. However, for experiments with large uncertainties or ambiguities in detector hit positions, or in experiments with high detector noise, low detector efficiencies, and high track multiplicity, track reconstruction can be challenging and computationally expensive \cite{SGUAZZONI20162497,Langenberg_2014}.

The computational burden on track reconstruction can be decreased by using pattern matching algorithms to prune data of 
noise and help discriminate signals that potentially correspond to particle tracks of interest from those produced by background processes. All of the hits that are produced from a single physics event constitute a pattern. This can be modeled as a binary array by dividing the detector into discrete positional segments and recording if a hit did or did not occur in each segment. Pattern matching algorithms identify potential track candidates by comparing the observed pattern from an event to a library of pre-computed signal patterns matching events of interest. The library of patterns is created using a simulation of the detector or through analysis of previous data. 

For detectors with a large number of discrete segments, the pattern libraries become equally large and yield a corresponding increase in the time needed to search the library. The search time can be decreased by organizing the library patterns into a tree structure of increasing resolution \cite{dell1990}. However, if the pattern from the event includes signals due to detector noise or has missing data due to detector inefficiencies, then the sought-after pattern will be distinct from the corresponding library pattern by one or more bits. The tree search algorithm's speed and effectiveness are therefore sensitive to the amount of detector noise, background processes, and detector efficiencies present, in addition to the detector's spatial resolution.  

In addition to pruning data before track reconstruction, pattern matching methods are implemented in hardware to decrease the amount of data recorded directly from an experiment. This is critically important for high luminosity experiments, in which the volume of recorded data must be filtered to remain manageable. Using pattern matching as a trigger for data acquisition requires  efficient algorithms in both time and storage.



Over the past few years, there have been several efforts to complete or improve the track recognition and reconstruction process using quantum computing. Track reconstruction efforts include using quantum annealing for pattern recognition when it is converted into a quadratic binary optimization problem (QUBO) \cite{bapst2020pattern}, and using gate-based quantum computers to complete quantum associative memory pattern recall \cite{shapoval2019quantum}. The research on using quantum computing for track reconstruction include using quantum annealing for optimization tasks related to charged particle track clustering \cite{das2019track} and track reconstruction via the Denby-Peterson method \cite{zlokapa2019charged}.

 Here we address the development of pattern matching for track recognition based on the principles of adiabatic quantum optimization. Our first approach utilizes an associative memory model (AMM), which represents a supervised learning model for pattern matching given incomplete or incorrect information~\cite{hopfield1982neural}. The second approach relies on content-addressable memory (CAM), an associative memory structure in which key-value (label-instance) data is recalled based on the value as opposed to the key~\cite{hopfield1982neural}. We demonstrate how both AMM and CAM may be used to implement track reconstruction based on binary detector models and quantum annealing (QA); thus, leading to a quantum AMM (QAMM) and quantum CAM (QCAM) approach to pattern matching for track recognition. QA is an computational method that exploits quantum mechanical phenomena to find the global minimum of an objective function~\cite{santoro:2002qa,johnson:2011dwave,mcgeoch:2014qa,albash:2018aqc}. Previous studies have shown that both AMM \cite{seddiqi2014adiabatic,santra2017ising} and CAM \cite{schrock2017recall} can be performed via QA. Potential computational advantages offered by QA-based methods have been suggested in the form of improvements to the theoretical learning capacity of the model. A theoretical study of QAMM argues that quantum tunneling may bring about exponential improvements in learning capacity over classical AMMs\cite{santra2017ising}. While experimental investigations of QAMM and QCAM have provided insight into hardware and algorithmic performance, these studies have been limited to artificial data sets based on orthogonal, non-orthogonal, and random patterns. Here, we broaden the scope to consider realistic datasets and examine the utility of this quantum approach to a real world application.

Alongside the development and demonstration of these ideas, we evaluate the the potential advantages of future particle track recognition methods that incorporate quantum computing techniques.  Our assessment tests the feasibility of using QAMM and QCAM for pattern matching in terms of the computational time required as well as the potential for maintaining or improving reconstruction performance.  Our approach characterizes and optimizes QAMM and QCAM performance for the practical application of track reconstruction using experimental QA hardware.

 
\section{Quantum Associative Memory}

\subsection{Quantum Annealing}
\label{subsec:aqo}
The pattern recognition problem is addressed by utilizing associative memory models in conjunction with QA, a finite-temperature, non-universal variant of adiabatic quantum optimization (AQO). In the ideal limit, AQO seeks to find the global minimum of an objective function by expressing that function as a Hamiltonian whose lowest energy eigenstate (i.e., ground state) defines a valid solution to the encoded optimization problem. The AQO algorithm defines the objective Hamiltonian as part of a time-dependent evolution that is ideally described by Schrodinger's equation. In AQO, the system is initialized in an easily prepared ground state. The Hamiltonian describing this initial state is deformed to the objective Hamiltonian. Provided this interpolation between Hamiltonians is sufficiently slow (i.e., in accordance with the adiabatic theorem of quantum mechanics~\cite{jansen2007:at,lidar2009:at}), the quantum system remains in the ground state with high probability at the end of the evolution~\cite{albash:2018aqc}. Given the direct equivalence between the ground state of the objective Hamiltonian and the computational solution, the time evolved state of the system at the end of the AQO corresponds directly to the solution to the optimization problem. 

The AQO algorithm is defined by the Hamiltonian
\begin{equation}
    H(t) = A(t) H_I + B(t) H_O,
\label{eq:H-ad}
\end{equation}
where $H_I$ and $H_O$ denote the initial and objective Hamiltonian, respectively. The annealing schedules $A(t)$ and $B(t)$ define the control of the evolution. For a given computational time $T$, the schedules are ideally chosen such that $A(0)=B(T)=1$ and $A(T)=B(0)=0$. The initial Hamiltonian is typically defined as
\begin{equation}
    H_I = -\sum^{N}_{j=1} X_j,
\end{equation}
where $X_j$ represents a tensor product of $N$ operators where all are the identity except for the $j$th operator that is given by the Pauli spin-1/2 $X$ operator
\begin{equation}
    X = \left(
    \begin{array}{cc}
       0  & 1\\
        1 & 0
    \end{array}
    \right).
\end{equation}
The initial Hamiltonian possesses a unique ground state corresponding to a uniform superposition of all possible computational basis states. The objective Hamiltonian is typically given by an Ising Hamiltonian composed of 2-local interactions of the form $Z_iZ_j$ and local bias terms proportional to $Z_i$. The operator $Z_i$ follows a similar structure to $X_i$, except that the $i$th operator in the product is given by 
\begin{equation}
    Z = \left(
    \begin{array}{cc}
      1  & 0\\
        0 & -1
    \end{array}
    \right).
\end{equation}
The AQO algorithm is implemented by evolving $H(t)$ from $t=0$ to $t=T$ starting from the equal superposition state and subsequently evolving to the ground state of $H_O$, as specified by the objective function.

We cast the associative memory recall problem as an AQO problem and solve it via QA. As a computational heuristic that does not strictly enforce adiabaticity, QA enables physical realizations of AQO on systems that operate at finite temperature and potentially under noisy conditions. The D-Wave processor exemplifies a physical implementation of QA that we leverage here for associative memory recall. As we will discuss below, by carefully selecting the interaction strengths and local biases of $H_O$, we embed the recall problem into an AQO problem that can be solved on the D-Wave quantum annealer.


\subsection{Quantum Associative Memory Model}
AMMs are supervised learning models that can be utilized to perform pattern matching given incomplete or incorrect information. Artificial neural networks designed to implement AMM paradigms, such as the Hopfield network \cite{hopfield1982neural}, are trained on a set of patterns $\mathcal{S}=\{\xi^{(\mu)}\}^{p}_{\mu=1}$ based on a particular learning rule. After encoding, the AMM can be exposed to a probe pattern $\chi$ that may or may not constitute a pattern in $\mathcal{S}$. The AMM performs pattern matching (or recall) by identifying the encoded pattern that obtains the greatest overlap with $\chi$; see Fig.~\ref{fig:qamm-qcam}.

AMMs can be cast as an AQO problem where the ground state solution to the objective function corresponds to the recall pattern. This QAMM is implemented via the objective Hamiltonian
\begin{equation}
    H_O = -\sum^{N}_{i,j=1} W_{ij} Z_i Z_j - \theta \sum^{N}_{i=1} \chi_i Z_i,
\end{equation}
where $W_{ij}$ is defined by the learning rule, and $\theta$ is the bias parameter. The parameter $N$ identifies the number of qubits, or equivalently, the length of the patterns defining the problem. The first term in the objective Hamiltonian defines a degenerate subspace whose ground states correspond to the encoded patterns $\{\ket{\xi^{(\mu)}}\}^{M}_{\mu=1}$. The second term breaks this degeneracy and biases the energy landscape to form a unique ground state or reduced subset of degenerate ground states that maximally overlap with the probe state $\ket{\chi}$. While there is potential for over biasing, optimal selection of $\theta$ can be performed by cross-validation or by utilizing previously developed bounds~\cite{Du:2019book}.

\subsection{Quantum Content Addressable Memory Model}
\begin{figure}[t]
\centering
\includegraphics[width=\linewidth]{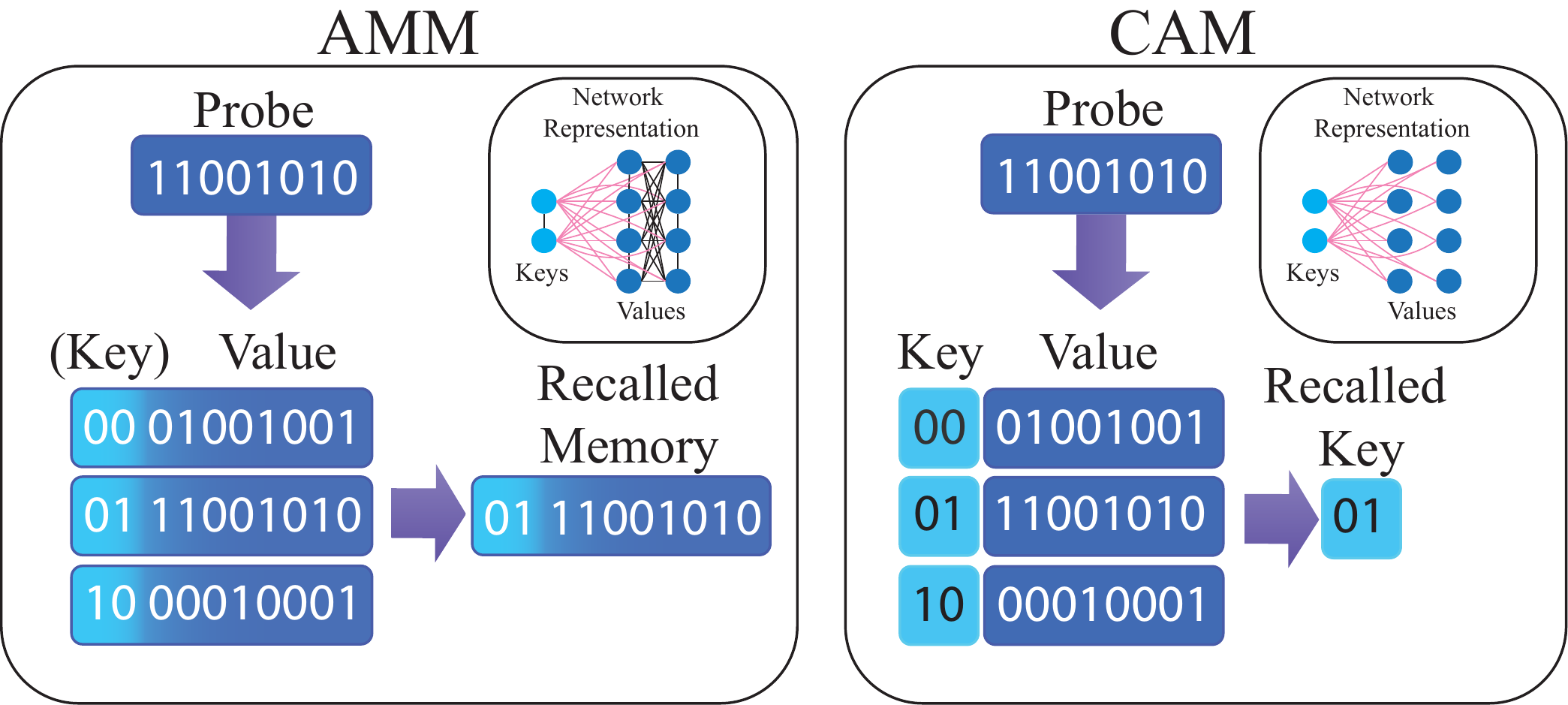}
\caption{Schematic comparison between AMM and CAM. Both models seek to recall patterns most closely matching the probe. Probe patterns are set by the values only. Training elements for AMM models may include keys (labels); however, unlike CAM, they are not partitioned in the learning rule. As a result, key-key and value-value interference is allowed in the AMM model. This distinction is conveyed by the network representations for AMM and CAM.}
\label{fig:qamm-qcam}
\end{figure}

CAM is an associative memory model that stores a relationship between a key $k$ and a value $v$, and recalls this association by retrieving $k$ when given $v$ or the closest $k$ when given an imperfect or approximate representation of $v$. CAM is distinct from alternative memory models, such as random access memory (RAM) in that CAM focuses on recalling a key given a value rather than recalling a value given a key; see Fig.~\ref{fig:qamm-qcam}. This distinction will become important here where keys denote the classification labels and CAM is used to perform binary classification.

CAM recall relies on a training set comprised of  key and value pairs. Hence, the encoded pattern set can be expressed as 
\begin{equation}
    \mathcal{S}=\{\xi^{(\mu)}=(k^{(\mu)}, v^{(\mu)})\}^{p}_{\mu=1}.
\end{equation}
Each pattern contains $N=K+V$ bits, where the key contains $K$ bits and the value is of length $V$. For the pattern matching problem considered here, the key bits represent classification labels and the values denote particle track patterns.

The CAM recall problem is cast as an AQO problem via the Hamiltonian
\begin{equation}
    H_O = -\sum^{N}_{i,j=1} W_{ij} Z_i Z_j - \theta \sum^{V}_{i=1} v^{(\mu^\prime)}_i Z_i,
\label{eq:Hp}
\end{equation}
where $W_{ij}$ is the weight matrix that encodes the pattern set $\mathcal{S}$ and $v^{(\mu^\prime)}$ designates the value of the probe pattern. As in the case of QAMMs, this quantum CAM (QCAM) Hamiltonian possesses a set of degenerate ground states associated with the encoded patterns. The second term in $H_O$ breaks the degeneracy and biases the energy landscape towards the encoded states that have the highest overlap with the probe state $\ket{v^{(\mu^\prime)}}$. This set of biased states defines the solution space for the AQO and thereby identifies the keys (classification labels) and values (patterns) that most closely resemble the probe pattern.

\subsection{Projection Rule Learning}
QAMM and QCAM recall performance is highly dependent upon the underlying learning rule used to train the model. The learning rule is a deterministic mapping that encodes the set of possible input patterns into the real-valued weight matrix $W$. Thus, in the case of AQO, the learning rule defines the interaction strengths between qubits required to specify the objective Hamiltonian.

While there are several possible choices for learning rules, we will focus on projection learning rules due to the characteristics of the data sets encountered in this study. Alternative learning rules, such as Hebb's rule typically perform best when the training set elements are orthogonal ~\cite{hebb:1949book}. Hebb's rule maps non-orthogonal patterns into overlapping projections, creating interference that can lead to inaccurate pattern recall. 

Projection rules are expected to be an improvement on Hebb's rule because they are designed to decorrelate the pattern set and minimize interference as part of the construction of $W$~\cite{personnaz:1986rule, kanter:1987rule}. The projection rule used in this study is defined by
\begin{equation}
    W_{ij} = \frac{1}{N}\sum^{p}_{\mu,\mu^\prime=1}\xi^{(\mu)}_i C^{-1}_{\mu\mu^\prime}\xi^{(\mu^\prime)}_j,
\end{equation}
where $C_{\mu\mu^\prime}=\frac{1}{N}\sum^N_{k=1}\xi^{(\mu)}_k\xi^{(\mu^\prime)}_k$ is the covariance matrix and $C^{-1}$ is the inverse of $C$. The theoretical learning capacity, i.e., the maximum number of patterns that can be encoded while still achieving perfect recall, is $N$ for orthogonal patterns and approximately $N/2$ for interfering patterns.

The projection rule will be used to train the QAMM, while a bipartite variation of the rule will be used to encode patterns into the QCAM model~\cite{schrock2017recall}. In principle, the projection rule can induce a fully connected graph between qubits. However, when taking advantage of the bipartite structure of the key and value relationship in QCAM, the connectivity graph can be reduced to a bipartite graph. The bipartite projection rule is obtained by zeroing out the elements of $W$ that correspond to key-key and value-value interference. The resulting weight matrix focuses on key-value interactions.

\section{Particle Tracks Dataset}
This paper covers several experiments aimed at characterizing the performance of QAMM and QCAM recall for pattern matching.  The performance of QAMM and QCAM recall is dependent on the number of bits in each pattern, the number of patterns in the pattern library, and the amount of incomplete and imperfect information.  The number of bits in each pattern (denoted as $V$) is equal to the number of segments in the detector, and $p$ represents the number of patterns in the library.  The amount of incomplete and imperfect information corresponds to the noise ($\gamma$), and efficiency levels ($\eta$) of the detector which can change one or more bits in a pattern from the patterns in the library, creating faulty patterns.  To develop both the pattern libraries and the probe patterns (those which we are trying to recall from the library) a flexible simulation of a detector was used.  Using a simulation, as opposed to experimental data, provided control over the QAMM and QCAM dependencies, such as $V$, $p$, $\eta$, and $\gamma$.

\begin{figure}[t]
\centering
\includegraphics[width=.75\linewidth]{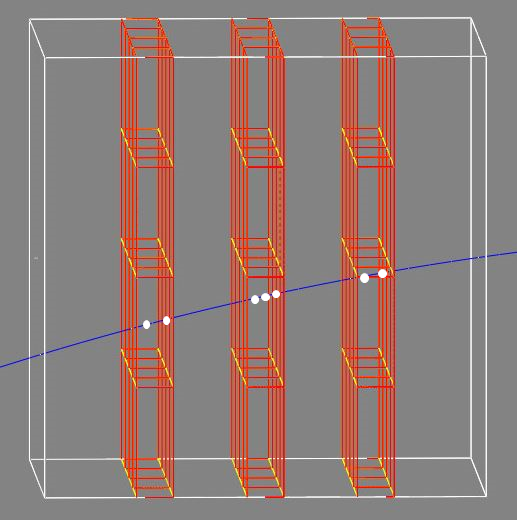}
\caption{
A diagram of the simulated detector with three parallel planes. Each plane displayed here is segmented into a 4 x 4 configuration. The blue line represents the track of an electron traversing the detector.}
\label{fig:toymodel}
\end{figure}

The detector simulation was developed using the Geant4 simulation and modeling toolkit \cite{geant4} and consisted of three parallel detector planes. The detector planes were segmented, as described in the introduction.  Electrons and positrons with energies of 0.5 GeV were initialized to the left of the detector and traversed the detector at angles close to perpendicular to the detector planes so the particles intersected with each plane exactly once.  The detector was placed in a uniform magnetic field (0.2 T) to allow for the particles to curve slightly as they traversed the detector. A three-dimensional diagram of the detector with one track is shown in Figure \ref{fig:toymodel}.

The lowest granularity considered a 24-segment detector by dividing each detector plane into eight segments (two by four) and the highest granularity was a 54-segment detector by dividing each plane into sixteen segments (three by six). Additionally, 30-segment, 36-segment, 42-segment, and 48-segment detectors were considered.

To create patterns from the simulated data, the segments in the detector were mapped to a bit set (an array of Boolean elements) of size $V$. To create a signal pattern, the corresponding bit in the bit set was set to 1 for all detector segments that registered a hit.  All other bits, that correspond to detector segments where no hit was registered, were set to 0.

The signal patterns were created by taking the particle track parameters of the patterns in the library, such as the particle identification, initialization location, and the four-momentum vector, and re-initializing the particles in the simulation, but with varying $\eta$ and $\gamma$.  To create the noisy signal patterns, the patterns were altered such that there was a $\gamma$ possibility of each 0-set bit being flipped to 1.  Signal patterns with efficiencies less than 100\% were created by taking patterns and randomly changing each “hit” bit to zero with a probability equal to $\eta$.  The efficiencies considered were $\eta = [1, 0.98, 0.96, 0.94, 0.92]$ and the random detector noise levels considered were $\gamma = [0, 0.02, 0.04, 0.06, 0.08]$.  We isolate the dependence of recall performance on $\eta$ and $\gamma$ by varying them independently.

Background patterns were also created.  These are patterns that do not correspond to a track, but are instead created by taking an bit set of zeros of size $V$ and randomly setting bits to 1 with a 15\% probability. The background patterns were compared to the signal patterns to make sure that they were at least 1 bit different to any signal patterns.

The pattern libraries were created one of two ways, depending on the type of experiment being run. For the first method, the pattern libraries were created by selecting $p$ unique perfect ($\eta = 1$ and $\gamma = 0$) signal patterns.  The second method, the pattern libraries included both unique perfect signal patterns and background patterns.  The choice in library ultimately depends upon the classifier selection. We will elaborate on this further in \ref{subsec:classifiers}.

Current hardware significantly limits both the size of the patterns that can be recalled and the number of patterns that can be stored in the library.  Because of this, QAMM and QCAM recall cannot be directly applied to pattern recognition for most experimental datasets.  To create suitable datasets given the current hardware constraints, a Hough-Transform can be used to subdivide a large experimental dataset into smaller datasets, both in $p$ and $v$. This method is described in detail in Section 5.4.

\section{Hardware Platform and Methods}
This study utilizes the D-Wave 2000Q-6 processor which implements QA using 2048 superconducting flux qubits. The device topology follows the Chimera architecture, with 16 cells composed of 8 qubits each. Each qubit is permitted approximately 6 tunable interactions, allowing for a total of 64 fully connected logical qubits. Each logical qubit may be composed of physical qubits ferromagnetically coupled to each other. Logical qubit chains are required when interactions between non-neighboring qubits are desired and vary in size depending upon problem specification and embedding procedure. Here, we utilize clique embedding to embed the recall problem on the D-Wave processor. The embeddings are fixed for all recall models and noise parameter, however, varying depending upon detector size.

Qubit couplings are scaled to satisfy conditions on intrachain couplings. The weights matrix $W$ defines the coupling strengths between qubits, and it scales as $1/N$. As the problem size increases, the coupling strengths between qubits can become smaller than the allowed precision (typically 4-5 bits) of the device. Coupling strengths are rescaled by factor of $\frac{3}{4W_{max}}$ in accordance with previous studies and preliminary simulations. $W_{max}$ is the maximum value of the weights matrix and the additional factor is chosen to ensure that intrachain ferromagnetic coupling strengths for logical qubits remain dominant~\cite{schrock2017recall}. Note that this scale factor is chosen for QAMM and QCAM models, and generally conveys favorable recall performance for all models and noise parameters considered.

Each QA experiment performed on the D-Wave processor outputs state configurations and associated energy values. Each state configuration identifies the state of each qubit at the end of the QA using a bipolar representation. Due to the stochastic nature of QA, it is typical to perform multiple annealing runs to collect statistics for each optimization problem; thus, the result of QA commonly includes multiple state configurations and energies corresponding to candidate solutions to the underlying optimization problem. Here, we perform each recall experiment using 100 annealing runs for an annealing time of $T=10\,\mu$s. Additional annealing times were examined; however, recall performance did not appear to vary significantly with this parameter. After performing QA, resulting embedded solution states are converted to an umembedded solution using majority vote to resolve broken logical chains. Umembedded solutions are used to calculate the average recall performance for a particular probe pattern based on all 100 samples.

\section{Results}
We apply QAMM and QCAM to the problem of distinguishing between signal and background patterns. Below, we first explore different approaches for defining a classifier and then examine the performance of each memory model as a function of the detector size. We evaluate the performance of QAMM and QCAM when attempting to recall perfect encoded patterns, and then turn our attention to faulty probe patterns characterized by noise and inefficiency in the detector system. Lastly, we examine the effect of manipulating the QA control schedules by assessing recall performance using what is known as reverse annealing.

\begin{figure}[t]
\centering
\includegraphics[width=\columnwidth]{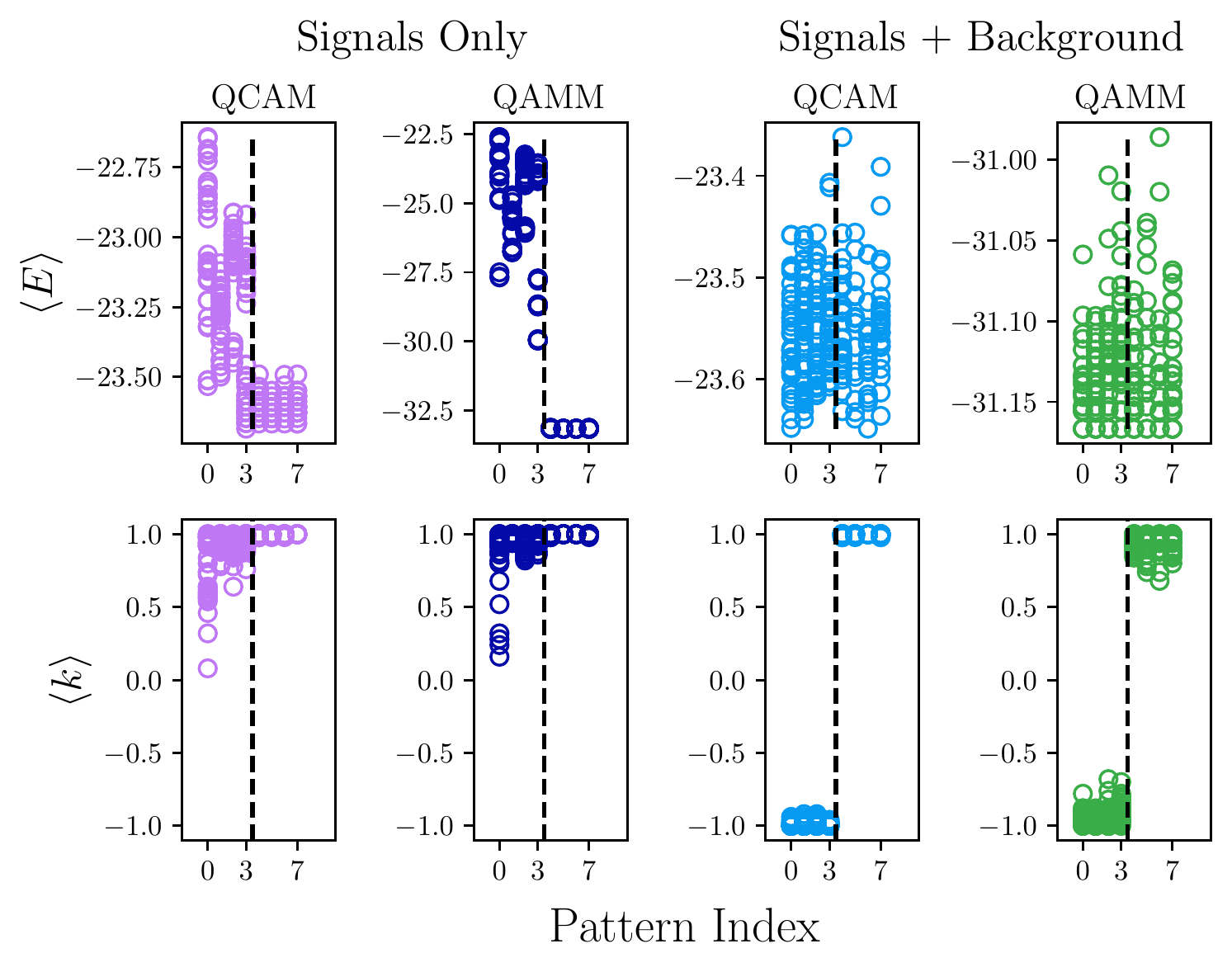}
\caption{Comparison between average energy and average key values for each recall method under different pattern encoding schemes. Columns one and two (from the left) convey signal pattern only encoding, while columns three and four use signals and background patterns. The dotted line in each subplot distinguishes results for background (left) and signal (right) probe patterns. Data shown for $V=24$ and a signal pattern density of $\alpha=1/6$.}
\label{fig:class_comp}
\end{figure}

\subsection{Classifiers}
\label{subsec:classifiers}
QA-based classification allows for more than one method for distinguishing between signal and background patterns. While the QCAM model explicitly defines keys to label training data, the QAMM model does not and therefore, unless the key is included in the encoded pattern, one requires an alternative approach for classification. Here, we consider an alternative classifier based on the energy of the solution states resulting from QA in addition to the key-based approach.

In both the QCAM and QAMM models, one can choose to designate the training set as either a combination of signal and background patterns or purely signal patterns. In the former, a key bit is required to distinguish probe patterns as signal or background since the ground state manifold of Eq.~(\ref{eq:Hp}) is composed of both pattern classes. In this case, examining the energy of the QA output states proves to be uninformative as a classifier. In contrast, when the training set is composed entirely of signal patterns, the opposite situation occurs. Background patterns are not encoded and therefore, the classification bits of the QA output states do not provide a mode for classification. Distinctions in energy, however, become more evident, implying that energy can be used as a mechanism for binary classification if the training set is chosen appropriately. Fig.~\ref{fig:class_comp} illustrates this behavior for the case of a 24-segment detector using 4 encoded signal patterns (signal only results) and a combined set of 4 signal and 4 background patterns (signal + background results). The results are compiled from 5 different training sets using 50 probes (25 signal and 25 background) for each training set. The vertical dotted line separates background recall (left) and signal recall (right). Average energy $\braket{E}$ and key $\braket{k}$ are calculated using 100 annealing runs for each probe experiment.

The recall bias parameter is fixed for this analysis and all subsequent comparisons. Preliminary studies on recall performance did not convey a strong dependence on the bias parameter after $\theta=0.1$, except for energy-based QCAM; see supplemental material. In order to effectively compare all recall models under near-optimal conditions, we select a bias that enables high recall performance for energy-based QCAM without the concern of over-biasing; hence, we choose $\theta=0.74$.

\begin{figure}[t]
    \centering
    \includegraphics[width=\columnwidth]{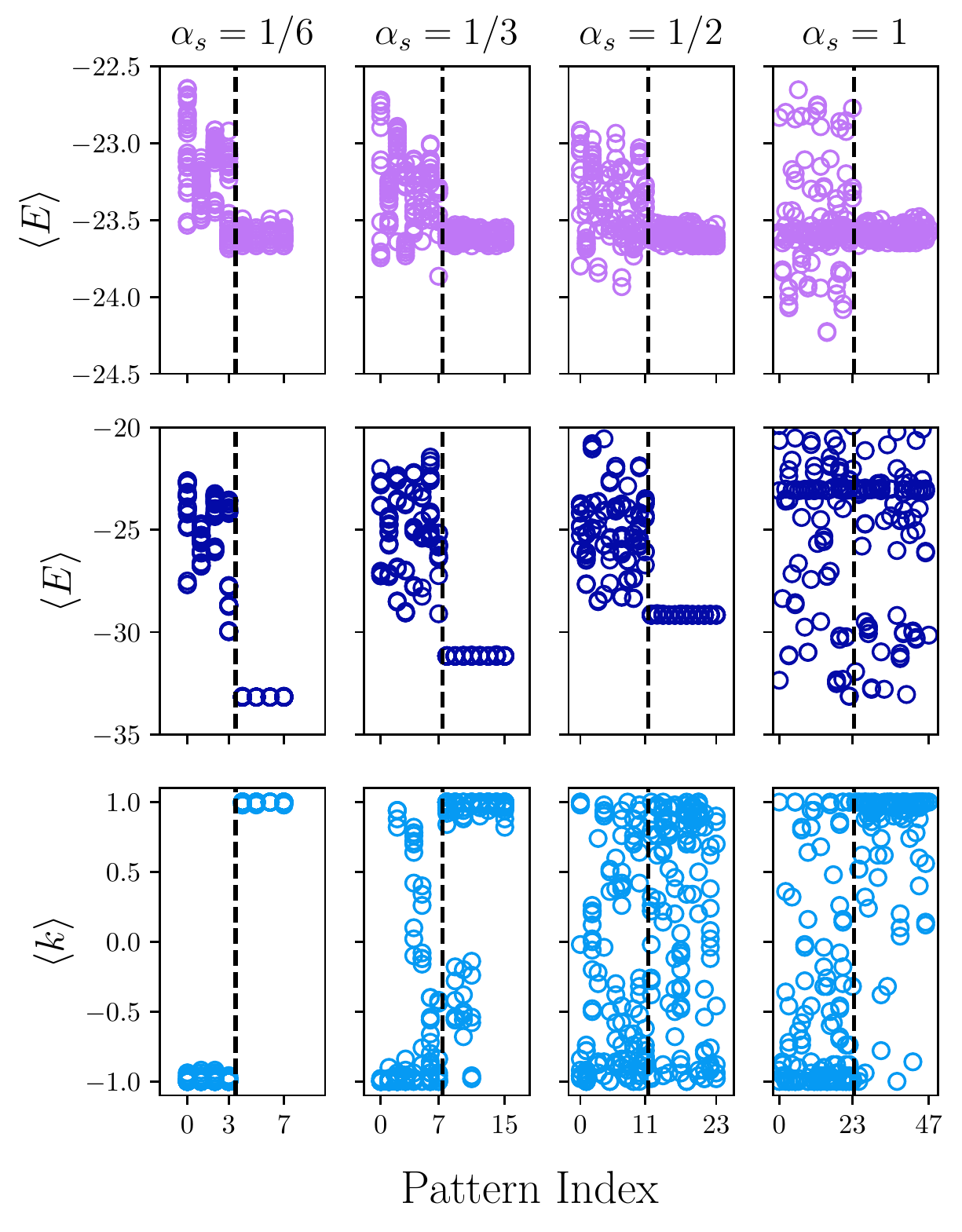}
    \caption{Comparison between signal-only QCAM energies (top row), signal-only QAMM energies (middle row) and signal + background QCAM key values (bottom row) as a function of encoded signal pattern density $\alpha_s$. Energy and key values are the result of 100 annealing runs collected from 5 unique training sets and 50 probe patterns for a $V=24$ cell detector. The dotted black line separates energy and key values from background (left) and signal (right) probe patterns. Results indicate signal and background distinguishability descreases with increasing $\alpha_s$.}
    \label{fig:vary_alpha}
\end{figure}

Classification accuracy in both the energy and key-based approaches is dependent upon the density of encoded patterns, where the density is defined as the number of signal patterns $p_s$ to the pattern (value) length $V$. While clear partitioning between signal and background probes is observed in Fig.~\ref{fig:class_comp}, these distinctions become less obvious as the number of encoded patterns increases; thus reducing classification accuracy. We display this behavior in Fig.~\ref{fig:vary_alpha} for QAMM energy (signal only encoding) and QCAM key values (signal and background encoding) for encoded signal pattern densities $\alpha_s=1/6,1/3, 1/2$, where $\alpha_s=p_s/V$. Equivalently, we define the background pattern density $\alpha_b=p_b/V$ with respect the number of encoded background patterns $p_b$. Here, we take $\alpha_b$ to be equal to $\alpha_s$. While one can consider lower density background pattern encodings, we find that $\alpha_s=\alpha_b$ typically results in the best recall performance. Therefore, all subsequent analyses using key-based classification center around an equivalent number of signal and background patterns in the training set.

The effect of value-value interference on pattern distinguishability can be observed from Fig.~\ref{fig:vary_alpha}. We include energy comparisons for QCAM with signal-only encoding in addition to signal-only QAMM in Fig.~\ref{fig:vary_alpha}. Distinctions between signal and background energies are generally similar between the two approaches, except for $\alpha_s=1$ where signal energies remain localized for QCAM and not for QAMM. We attribute this distinct behavior to value-value interference for QAMM and the lack thereof for QCAM. Note that while we show data for $\alpha_s=1$, this trend is generally observed as $\alpha_s$ increases and becomes more pronounced after $\alpha_s=2/3$.  Below, we will explore the effect of this localization on recall accuracy.

The behavior observed in Fig.~\ref{fig:vary_alpha} is consistent with classical associative memory recall theory, which imposes bounds on recall performance based on encoding pattern density and learning rule. Moreover, while optimal recall for quantum associative memory appears to be consistent with classical bounds for the projection learning rule used in this study, we observe interesting recurrences in performance that appear to be unique to the quantum approach. We elaborate on this unexpected behavior below.

\begin{figure*}[t]
\centering
\includegraphics[width=\textwidth]{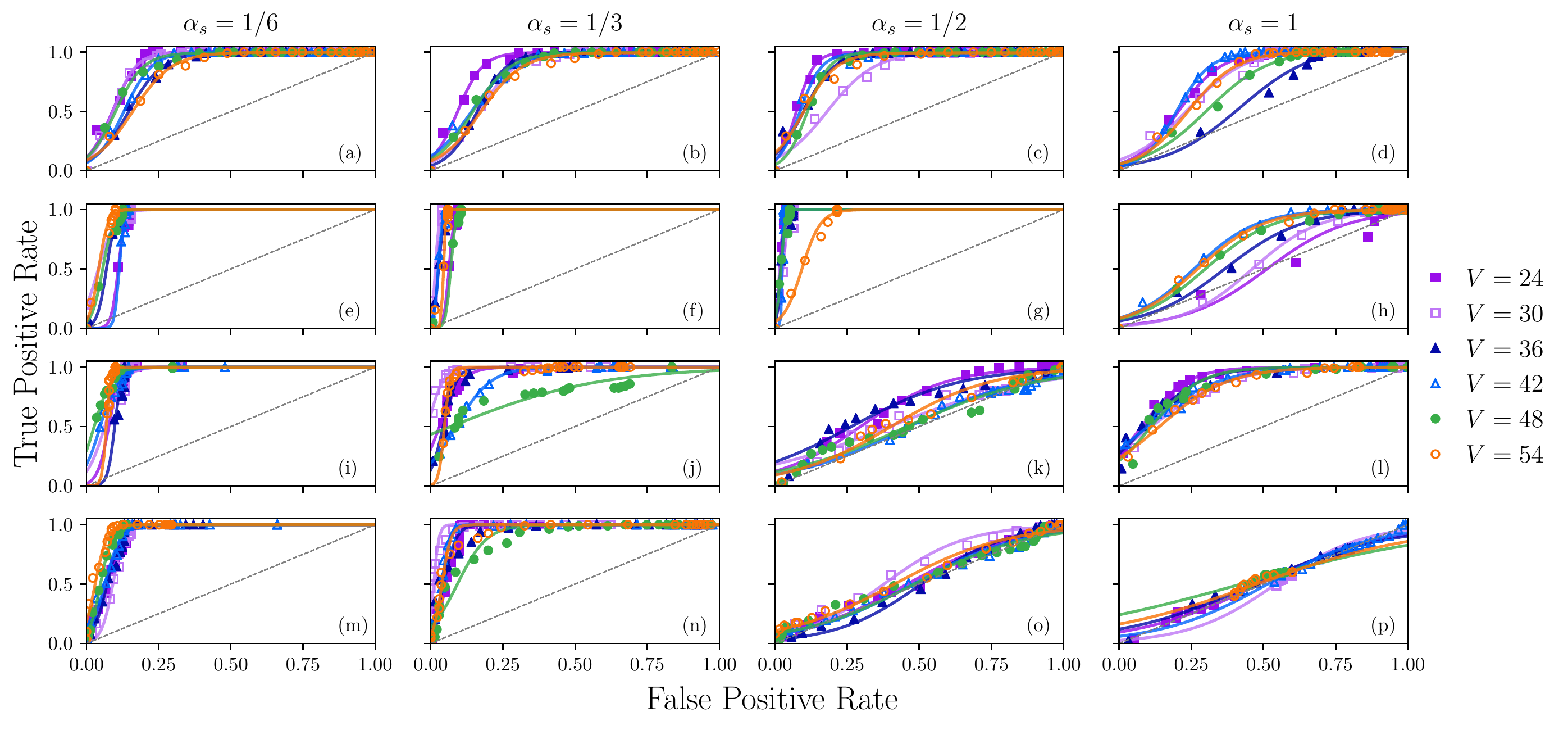}
\caption{Comparison between recall models as a function encoded signal pattern density $\alpha_s$ for varying detector size $V$. Models considered include: energy based QCAM [panels (a)-(d)], energy based QAMM [panels (e)-(h)], key-based QCAM [panels (i)-(l)], and key-based QAMM [panels (m)-(p)]. Each panel displays a ROC curve for each detector size, where data points are generated by varying the signal discrimination threshold. Data is averaged over 5 unique training sets and 50 unique probes (25 signal and 25 background patterns). Logistic fits (solid lines) are included to elucidate data trends. Data conveys that energy-based QAMM generally achieves the best recall performance for $\alpha_s<1$, while key-based QCAM performs best for large $\alpha_s\geq1$.}
\label{fig:enc_scaling}
\end{figure*}

\subsection{Noiseless Pattern Recall}
Utilizing both energy and key-based classification methods, we examine the performance of QCAM and QAMM recall as a function of detector size and encoded pattern density. Energy-based classification is based on a signal pattern energy range that is determined by performing encoded signal pattern recall on the QPU. The signal discrimination range is set by $\braket{E}\pm\beta\,\sigma_E$, where $\braket{E}$ is the mean of the encoded signal probe set and $\sigma_E$ is the standard deviation. The parameter $\beta$ is a real number that effectively controls the size of the discrimination range. A similar approach is used for key-based classification, where the standard deviation $\sigma_k$ is used to define a range for acceptable key values for signal classification.

The performance of each classifier and recall model is assessed via calculation of the true positive rate (TPR) and false-positive rate (FPR). TPR is defined as the ratio of true positive counts to the sum of true positives and false negatives. FPR is defined similarly as the ratio of false-positive counts to the sum of false positives and true negatives. Correct signal classification is defined as a true positive, while false positive classification denotes the classification of a background pattern as a signal. The definitions of true negative, and false negative follow accordingly. We display TPR and FPR data using receiver operator characteristic (ROC) curves. TPR and FPR data for each ROC curve is generated by varying the signal discrimination threshold via $\beta$. In this study, we consider $\beta\in[0,10]$; hence, representing acceptable signal tolerances as large as ten standard deviations from the mean.

Using the metrics defined above, we first examine QCAM and QAMM recall performance for both classifiers under the condition of noiseless pattern recall. Noiseless signal probe patterns are defined as signal patterns that perfectly match patterns encoded within the model. Noiseless background probe patterns are defined similarly with the caveat that background probes will only perfectly match patterns within the training set when performing key-based classification. We summarize the results for noiseless recall in Fig.~\ref{fig:enc_scaling} for varying detector size and encoded signal pattern density. ROC curves for energy-based classification for QCAM and QAMM are shown in panels (a)-(d) and (e)-(h), respectively. Panels (i)-(l) denote key-based QCAM performance, while panels (m)-(p) convey QAMM recall using key-based classification. Each column contains a fixed encoded pattern density, where each panel displays ROC curves for a variety of detector sizes. The data is generated from 5 unique training sets and 50 unique probe patterns (25 signal and 25 background). ROC curves are produced by varying $\beta$ over the range discussed above. Each data set includes a logistic fit to help guide the eye. Note that ROC curves favoring the top-left corner of each subplot denote optimal discriminators, while those tending towards the diagonal, black, dotted line designate models with the least discriminatory capabilities. In certain cases, for example, panel (o), the data drops below the diagonal, denoting reciprocated class discrimination. This behavior is likely due to the presence of spurious memories, which elaborate on below.

\begin{figure}[b!]
    \centering
    \includegraphics[width=1.0\columnwidth]{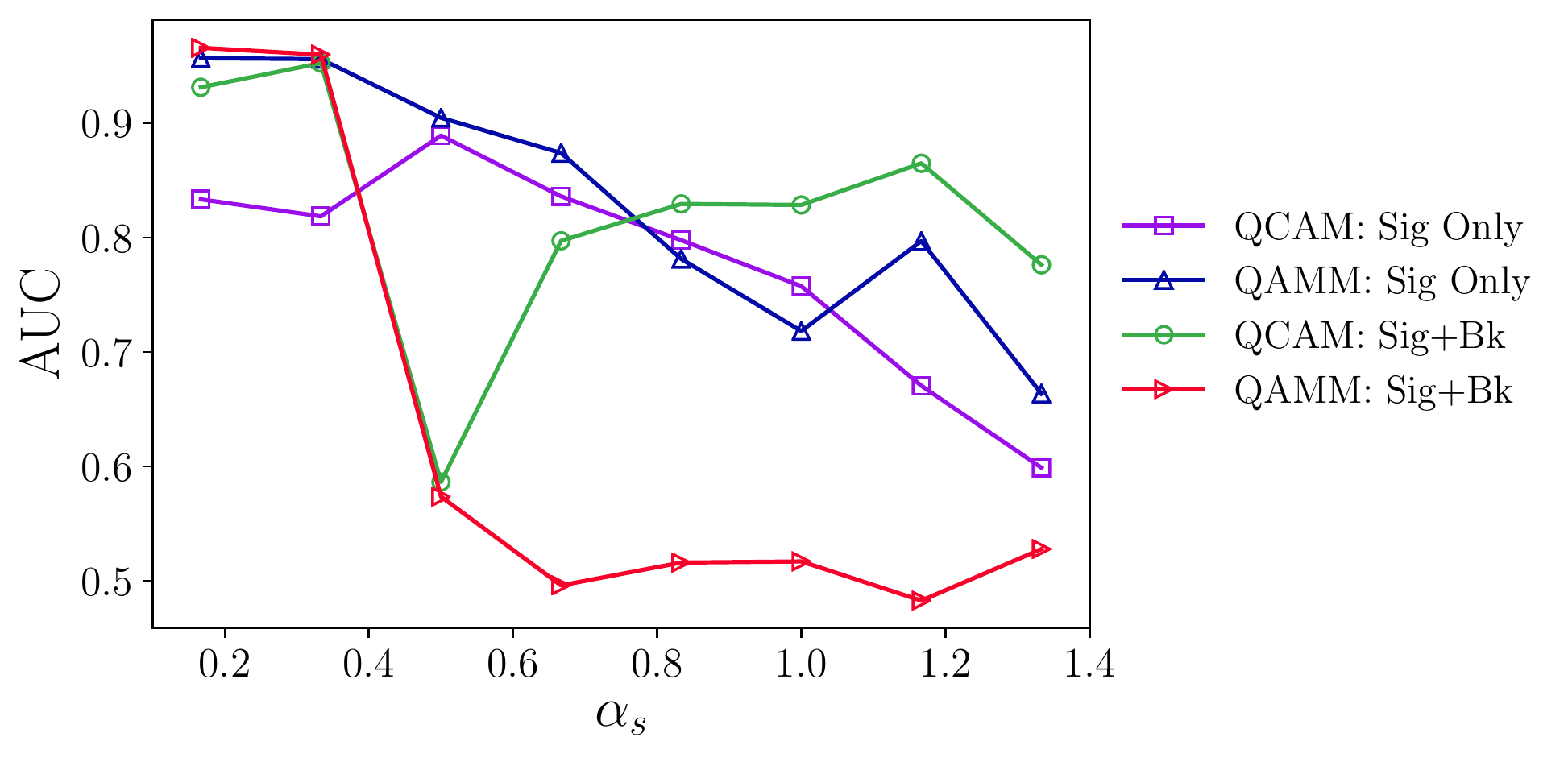}
    \caption{Comparison between recall models from the perspective of AUC as a function of encoded signal pattern density $\alpha_s$. Results are shown for a $V=54$ cell detector. Recall performance is nearly equivalent for all methods except energy-based QCAM up to $\alpha_s=1/3$. Energy-based QAMM provides the best signal discrimination up to $\alpha_s=5/6$. key-based QCAM dominates thereafter for the remaining $\alpha_s$ values considered in this study.}
    \label{fig:auc_encoded}
\end{figure}
\begin{figure*}[t!]
\centering
\includegraphics[width=1.0\textwidth]{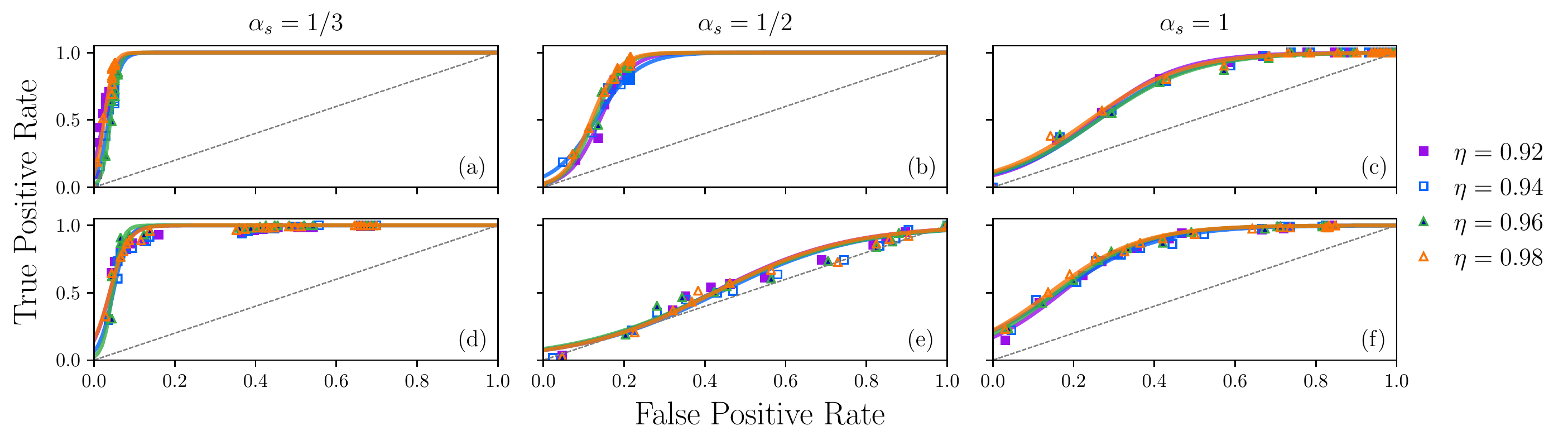}
\caption{Comparison between recall models for varying detector efficiency $\eta$. Data focuses on energy based QAMM [panels (a)-(c)] and key-based QCAM [panels (d)-(f)] with $V=54$ detector cells. Each panel includes ROC curves averaged over 5 unique training sets and 50 probe patterns composed of 25 noisy signal patterns and 25 background patterns. Logistic fits are included to show data trends. Recall models convey a relatively high degree of robust to detector inefficiency; see supplemental material for additional data.}
\label{fig:eff_scaling}
\end{figure*}

Our results indicate that energy-based QAMM is typically the best model for signal/background discrimination. While each model possess some reduction in performance with increasing encoded pattern density, energy-based QAMM maintains a large TPR and low FPR typically up to $\alpha_s=1/2$. Thereafter, a sharp transition in performance appears, where the model is no longer able to adequately discriminate between signal and background. Similar trends in performance as a function of $\alpha_s$ are observed for most of the studied models, where performance begins to substantially degrade between $\alpha_s=1/3$ and $\alpha_s=1/2$. According to the maximum theoretical learning capacity for the projection rule, one expects performance to quickly decline after the total (signal + background) encoded pattern density reaches $\alpha=1/2$.  In the case of energy-based classification, this behavior generally holds. Key-based classification appears to go beyond this limit, achieving consistently high recall performance up to $\alpha=2\alpha_s=2/3$. A previous theoretical study on QA-based associative memory argued that conventional learning capacity limits could be surpassed by QA due to the presence of quantum tunneling~\cite{santra2017ising}. It is unclear if we are observing evidence of such an advantage, however, the trends consistently appear for bias strengths above $\theta=0.1$ and all detector scenarios considered. Further investigation is require to determine if this observation is definitive experimental evidence of enhanced learning capacity brought about by quantum mechanical phenomena.

Although model performance tends to diminish with increasing $\alpha_s$, not all models exhibit the same behavior. In particular, a sharp decline in recall performance is observed at $\alpha_s=1/2$ for key-based QCAM followed by an immediate increase in performance observed at $\alpha_s=2/3$ that is generally maintained for the remaining $\alpha_s$ values considered in this study. Despite never achieving previous TPRs and FPRs found for $\alpha_s<1/2$, the increase in performance is sufficient to surpass all other methods and deem key-based QCAM as the best performer for $\alpha_s\gtrsim 1$. Such behavior has not been previously observed for QCAM and is likely due to the datasets considered. Whether or not such trends would be present for all non-orthogonal or non-random pattern sets, and perhaps what conditions on pattern overlap would be required to observe this phenomenon, is an open question.

Recall performance trends observed for all models are further explored in Fig.~\ref{fig:auc_encoded}, where the area under the ROC (AUC) is shown as a function of pattern density $\alpha_s$ for $V=54$. AUC provides an alternative perspective on signal/background discrimination, where perfect discrimination is denoted by unity and poor separability in the data is represented by AUC$=0.5$. Fig.~\ref{fig:auc_encoded} summarizes the typical trends in Fig.~\ref{fig:enc_scaling}, indicating nearly equivalent performance for most methods for small $\alpha_s$, transitions in performance (most notably for key-based QAMM and QCAM) after $\alpha_s=1/3$, and eventual shifts in the best-performing discriminator. Together, both perspectives convey that energy based QAMM is typically the best discriminator for small $\alpha_s$, while key-based QCAM tends to provide the best discrimination for large $\alpha_s$. The remaining methods either perform similarly or worse than the best performers.

\begin{figure*}[t]
\centering
\includegraphics[width=\textwidth]{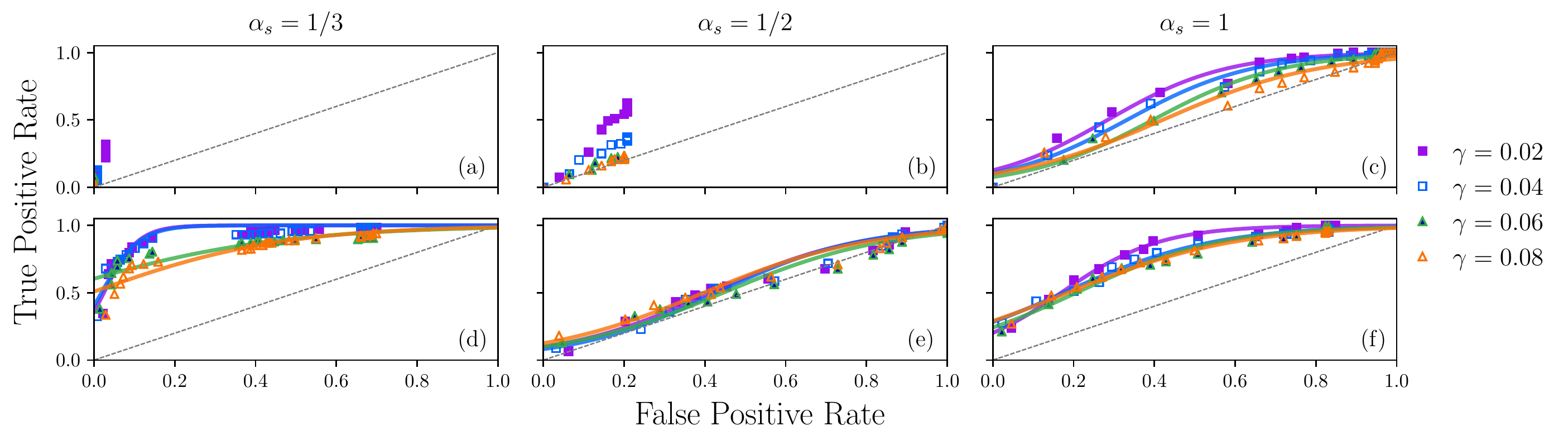}
\caption{Comparison between recall models for varying detector noise $\gamma$. Data focuses on energy based QAMM [panels (a)-(c)] and key-based QCAM [panels (d)-(f)] with $V=54$ detector cells. Each panel includes ROC curves averaged over 5 unique training sets and 50 probe patterns composed of 25 faulty signal patterns and 25 background patterns. Logistic fits are included to show data trends where applicable. Energy-based QAMM shows significant performance degradation with detector noise, while key-based QCAM remarkably exhibits considerable robustness to detector noise; see supplemental material for additional data.}
\label{fig:noise_scaling}
\end{figure*}

Distinctions between key-based QCAM and energy based QAMM for large $\alpha_s$ are likely due to encoded pattern interference. QAMM permits key-key, key-value, and value-value interference. In the case of binary classification, key-key interference is non-existent and therefore, value-value interference is the only form of ``self-interference" between encoded patterns. As the encoded pattern density increases, value-value interference transitions from being an asset to a detriment, eventually resulting in an inability of the model to effectively discriminate between signal and background. This trend is not limited to key-based classification, but also extends to energy-based classification, where increased value-value interference leads to amplified degeneracy between signal and background recall energies; see Fig.~\ref{fig:vary_alpha}. The unfavorable effect of value-value interference on energy-based recall performance can be clearly observed by comparing QCAM and QAMM in Fig.~\ref{fig:enc_scaling}(d) and (h), respectively. 

Despite both energy-based and key-based QCAM models suppressing value-value interference, there still appears to be an advantage to key-based recall for large $\alpha_s$. We attribute this behavior to the susceptibility of energy-based classification to spurious pattern recall. While the task of the learning rule is to create fixed points (local minima) at specific state-space locations, most learning rules naturally manifest additional fixed points at locations that do not represent an encoded pattern. These additional fixed points constitute unwanted spurious patterns that can lead to incorrect memory recall~\cite{athithan1997:sm, fan2008:amm}. Each fixed point is said to be an attractor with an associated basin of attraction. Ideally, the learning rule maintains larger basins of attractions for correct patterns and smaller basins for spurious patterns. However, even if such a condition can be achieved, as the encoded pattern density increases, so does the number of spurious patterns. Here, we observe the result of an increase in spurious pattern recall when the probe pattern is a background pattern. Examining Fig.~\ref{fig:vary_alpha}, we find that recall energy for background probes begins to localize around the recall energy expected for signal probes as the encoded pattern density increases. Since the background patterns are quite distinct from the encoded signal patterns, this localization in energy is likely due to an increased attraction towards spurious signal patterns with the same energy as correct signals. These experimental findings are consistent with previous studies of spurious patterns and basins of attraction in QA-based associative memory~\cite{santra2017ising}.

Recall error due to the presence of spurious patterns appears to be less prominent for key-based QCAM than its energy based counterpart. We believe this inherent robustness is simply due to key-based QCAM only relying on the state of a single qubit. In key-based QCAM, both signal and background patterns are encoded into the model. Spurious patterns are naturally created for both pattern types. If spurious background patterns closely resemble a signal probe, the model will mistakenly recall a spurious background pattern with a key value denoting recognition of a signal pattern. We suspect that this behavior affords an unexpected model robustness as the encoded pattern density increases; compare Fig.~\ref{fig:enc_scaling}(d) and (l).


\subsection{Faulty Pattern Recall}
Noiseless pattern classification provides a baseline for model performance in an ideal probe pattern scenario. However, to gain a more comprehensive understanding of the utility of each model for classifying realistic datasets, one must account for detector inefficiency and noise. For this reason, we now turn our attention to faulty pattern recall, where imperfect signal probe patterns are represented by encoded signal patterns with uniformly random bit flips with probabilities set by efficiency $\eta$ or noise strength $\gamma$. In order to isolate the effect of detector noise on pattern recall, $\eta$ and $\gamma$ are varied independently. Furthermore, we choose background probe patterns from the training set as in the noiseless recall case.

A strong dependence on detector inefficiency is not observed for the discrimination models we study. Utilizing the same training sets from the noiseless pattern recall examination, we impose detector inefficiencies on the 25 signal probe patterns for each of the five unique training sets. In Fig.~\ref{fig:eff_scaling}, we summarize the results of these experiments for $V=54$ and $\alpha_s=1/3,1/2,1$, and include additional data as supplemental material. We focus on the best performing recall models and therefore, display results for energy based QAMM in panels (a)-(c) and key-based QCAM in panels (d)-(f). Each subplot includes ROC curves for a range of $\eta$ values. Dependence on detector efficiency generally follows expectations: recall performance decreases with decreasing $\eta$. Significant variations in performance are not observed, indicating that our discrimination models are generally robust to detector inefficiency. Trends in performance as a function of pattern density are similar to the noiseless recall case, with only slight reductions in performance due to the inclusion of imperfect probes. Key-based QCAM recall performance again appears to decrease dramatically at $\alpha_s=1/2$ and improve thereafter. Overall, all models convey a similar degree of robustness to detector inefficiency.

In contrast, detector noise imparts a noticeable effect on model performance, dramatically reducing discrimination capabilities for certain recall models. In Fig.~\ref{fig:noise_scaling}, we show recall performance as a function of noise strength $\gamma$ for energy based QAMM [panels (a)-(c)] and key-based QCAM [panels (d)-(f)] for $\alpha_s=1/3,1/2,1$. Interestingly, QAMM performance reduces dramatically, nearly exhibiting a TPR of zero for $\alpha_s=1/3$ and indiscernible signal/background discrimination for larger pattern densities. Key-based QCAM, however, offers some robustness, still maintaining discrimination capabilities for $\alpha_s<1/2$ and $\alpha_s\geq 1$. This mild robustness of QCAM also appears for energy-based classification, indicating that it is likely due to the intrinsic structure of the QCAM model rather than the classification method. More concretely, suppressing self-interference between keys and values can lead to increased robustness in discrimination as probes become increasingly distinct from encoded patterns. This attractive attribute of the QCAM approach implies that it is likely a more appropriate candidate for signal/background pattern discrimination for real detector datasets. Additional data for energy-based QCAM classification and AUC as a function of $\alpha_s$ can be found in the supplement. 


\subsection{Faulty Pattern Recall via Reverse Annealing}
Thus far, our analysis has focused on quantum associative memory using QA with forward annealing control. Here, we exploit some of the D-WAVE QPU's advance control features and evaluate their potential ability to enhance faulty pattern recall. We focus on the best performing recall method under the most detrimental form of noise: key-based QCAM model in the presence of detector noise.

Reverse annealing has been shown to improve the performance of QA for certain applications~\cite{passarelli:2020rev, ohkuwa:2018rev,ikeda:2019rev,Chancellor_2017, venturelli:2019rev,rocutto:2020rev,grant:2020rev}. Namely, those applications in which the underlying optimization problems possesses local minima that are close in Hamming distance to global minima; therefore, local refinements improve the quality of solutions. While we observe that forward annealing solutions are reasonably close in Hamming distance to desired patterns, typically differing by 1-5 bits, it is difficult to determine to what extent one expects reverse annealing to improve recall performance. Hence, we must turn to an experimental investigation to gain insight into the potential advantages of reverse annealing. 

Forward annealing refers to the canonical QA protocol described in Sec.~\ref{subsec:aqo}, where the quantum system is evolved from the initial Hamiltonian to the problem Hamiltonian and then the state of the system is measured. In contrast, reverse annealing begins with the quantum system in a candidate classical state where the Hamiltonian of the system is defined by the problem Hamiltonian. The system is then driven in "reverse" by ramping down the $B(t)$ control schedule while simultaneously increasing the strength of the initial Hamiltonian. This process allows quantum fluctuations to increase and continue to contribute to the dynamics over a fixed time $t_{pause}$. After this selected time duration has expired, the system is driven according to the standard forward anneal and subsequently measured. 

In order to obtain candidate starting states for the reverse anneal, we perform a forward anneal; thus, producing a candidate recall pattern solution. The reverse anneal is performed 100 times, where the initial starting state is the candidate recall pattern and all subsequent starting states are the resulting state of the previous reverse anneal. For all experiments, we set the initial and final ramp duration to $1\mu$s and the intermediate pause time to $t_{pause}=10\mu$s. The initial ramp takes the system to a state described by $H(t^*)$, where the time $t^*$ is related to the normalized time parameter $s^*$ via $s^*=t^*/T$. Varying $s^*$ over a range of values, we examine the average TPR and average FPR to determine a favorable value $s^*=0.5$; see the supplemental material for more information. The remaining parameters (i.e., ramp duration and $t_{pause}$) were selected based on a brief evaluation of recall performance as a function of $t_{pause}$ for equal initial and final ramp times. A more thorough search for optimized reverse annealing parameters using e.g., closed-loop optimization techniques \cite{quiroz:2019cloaqc, pelofske2020:ctrl} is left for future work.

\begin{figure}[t]
    \centering
    \includegraphics[width=\columnwidth]{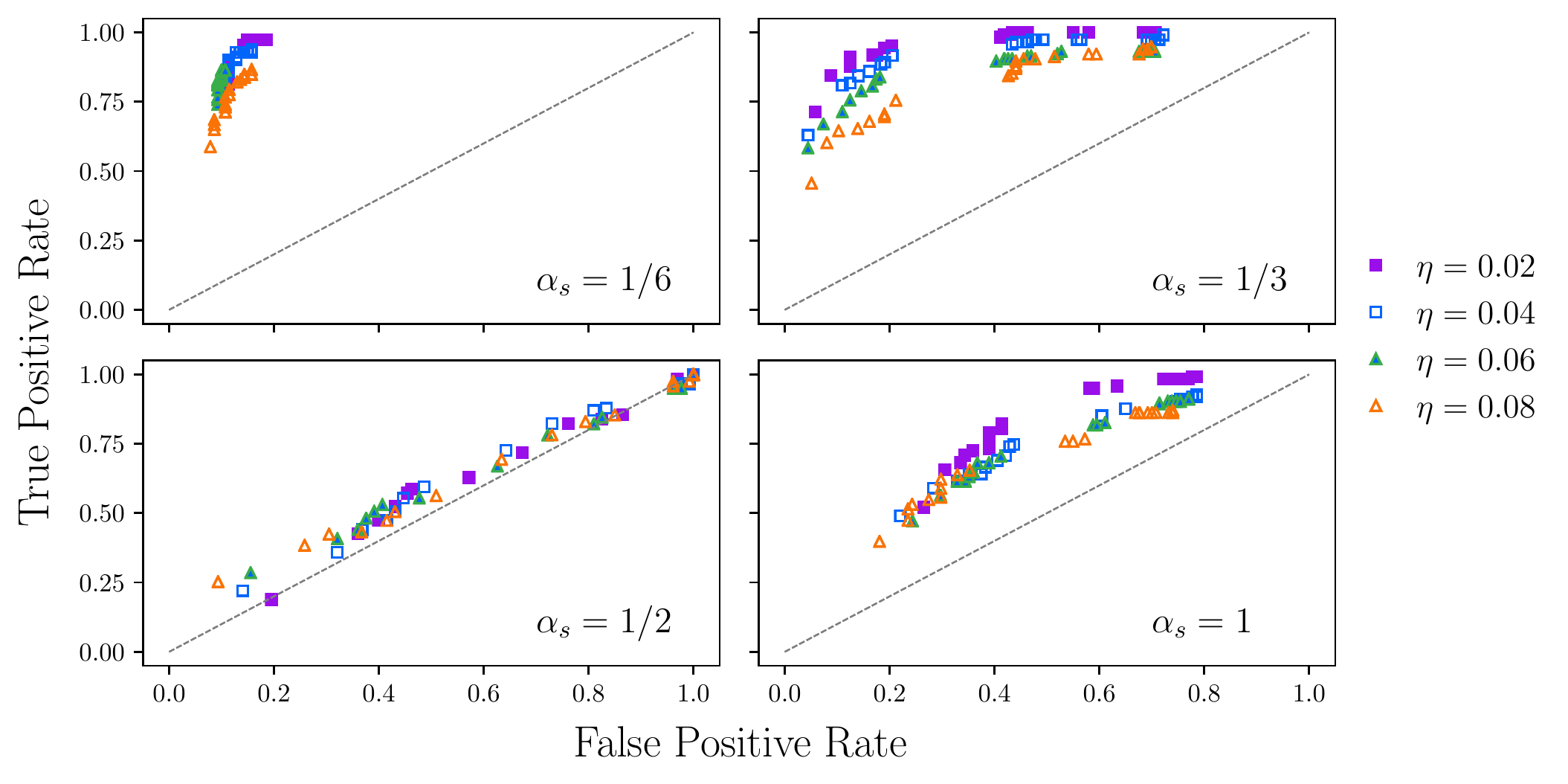}
    \caption{Recall performance as a function of detector noise for key-based QCAM using QA with reverse annealing control. Initial and final annealing quenches are $1\mu$s with an intermediate pause duration of $t_{pause}=10\mu$s. Reverse annealing leads to improved TPR at the cost of increased FPR.}
    \label{fig:rev_anneal}
\end{figure}

Reverse annealing improves TPR at the cost of a minor increase in FPR. In Fig.~\ref{fig:rev_anneal}, we show the results of reverse annealing recall for $V=54$ under a variety of encoded signal pattern densities and detector noise strengths. The most prominent improvement in TPR over forward annealing is found for $\alpha_s=1/6,1/3$. However, recall performance enhancements due to reverse annealing quickly diminish as pattern density increases.

Local refinements prove to be most beneficial for TPRs, suggesting that key-based QCAM is better at identifying faulty signals than background. Reverse annealing runs are seeded by solutions found by forward annealing runs. If the seed solution differs greatly from the desired pattern, local refinements in the solution will not improve its quality. We observe this behavior for FPRs, where $s^*$ values selected near the end of the annealing schedule typically result in the highest FPRs. On the other hand, FPR improve for $s^*$ near the beginning of the annealing schedule where quantum fluctuations are larger, indicating that forward annealing using background probes produces solutions that are trapped in deep local minima and require additional kinetic energy to properly traverse the objective landscape. TPRs are higher for $s^*$ near the end of the annealing schedule, and therefore, there exists an intermediate $s^*$ where both TPR and FPR can be simultaneously optimized. While we find $s^*=0.5$ to be a reasonable value here, it is likely that one can improve both quantities further by optimizing the forward and reverse annealing schedules.

\subsection{Extension to High-Energy Physics Collider Experiments}
The number of collisions in future collider experiments such as the High Luminosity LHC (HL-LHC), will increase by at least a factor of five. These conditions imply larger track multiplicities and will, therefore, complicate the problem of track reconstruction.

\begin{figure*}[t!]
\centering
\includegraphics[width=0.95\textwidth]{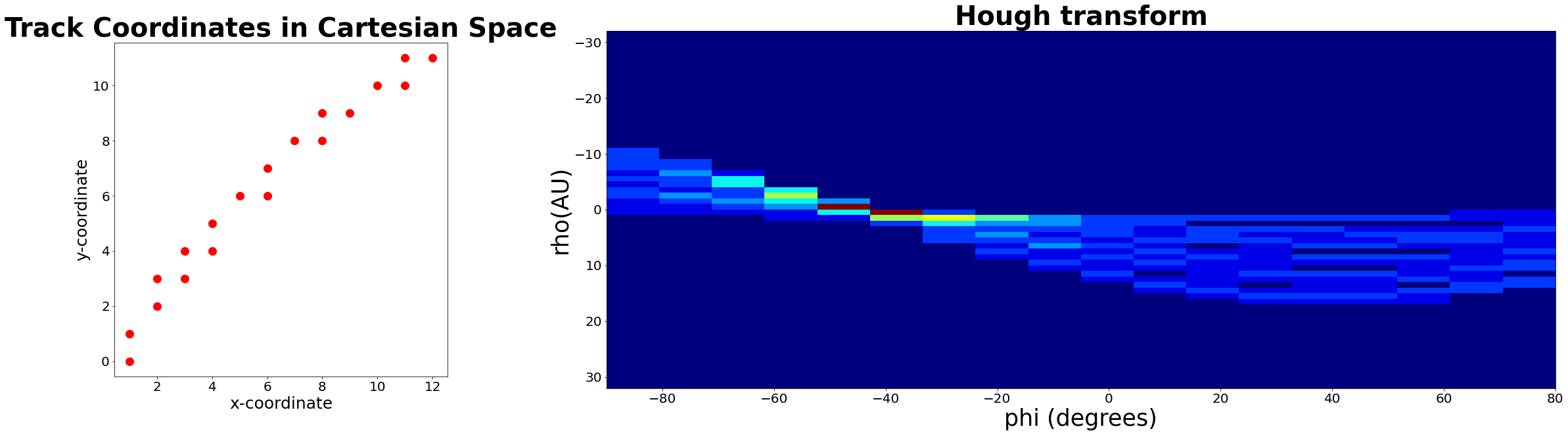}\\
\includegraphics[width=0.45\textwidth]{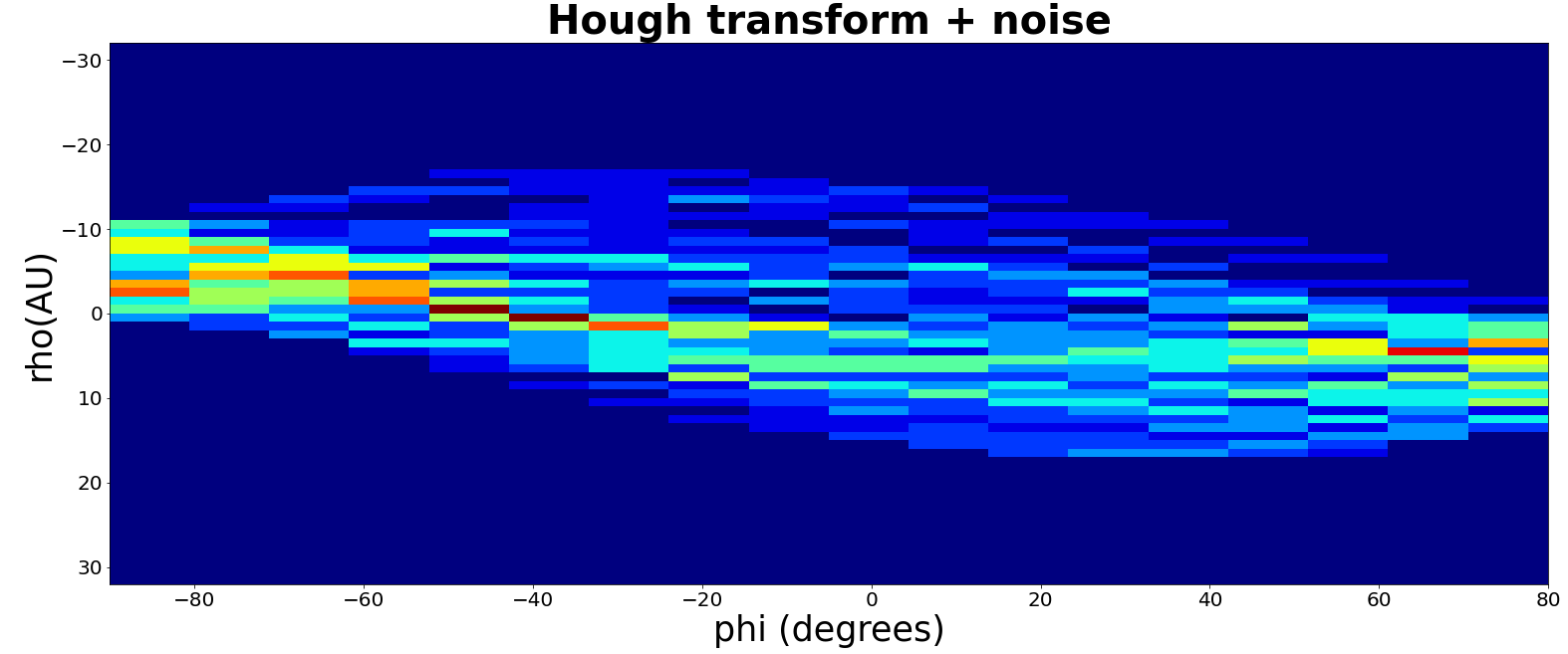}
\includegraphics[width=0.45\textwidth]{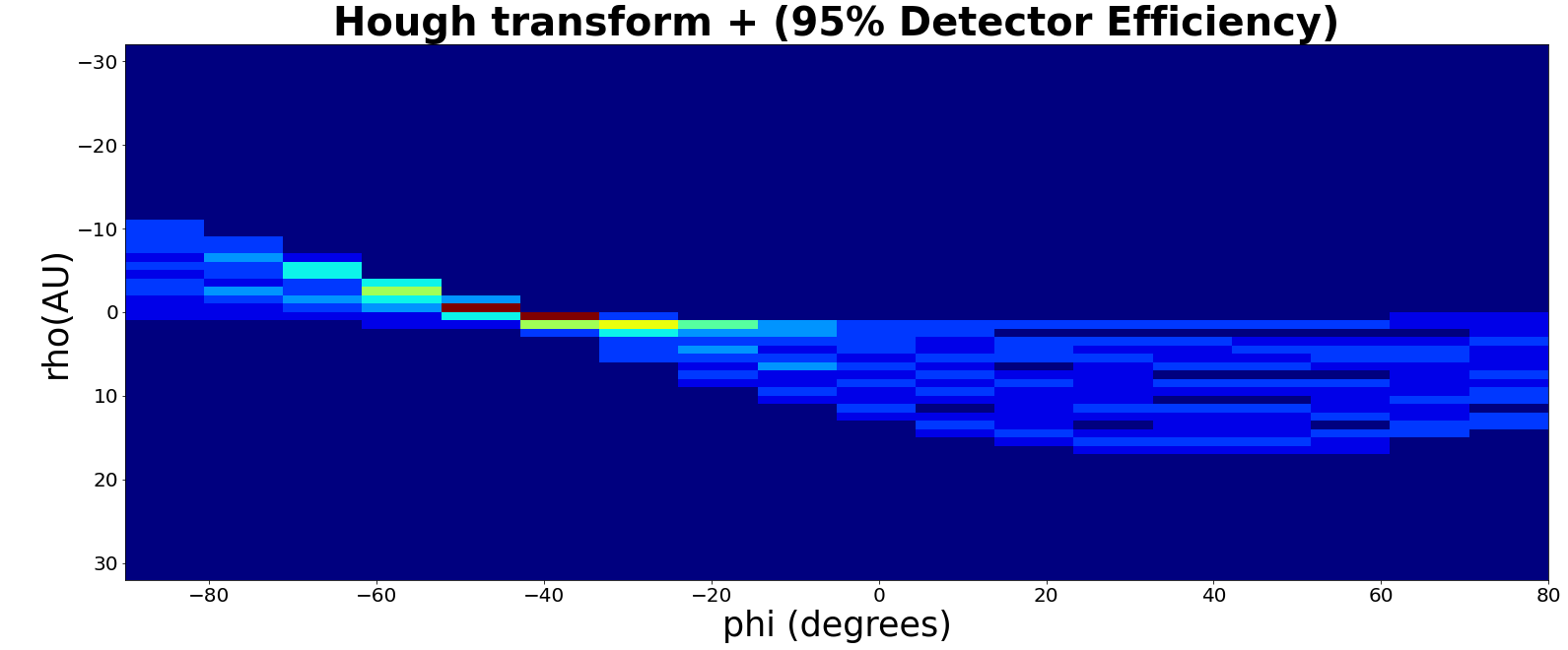}
\caption{The cartesian coordinates for detector hits corresponding to the track generated by a 0.5 GeV electron (top left). These points are then transformed into lines in Hough $\phi$ - $\rho$ space by using \ref{eq:htsp} (top right). The bin where these lines intersect correspond to the parameters of the track. The bottom plots show how the HT transform is affected by noise (left) and detector efficiency (right) for the same reconstructed track.}
\label{fig:httrack}
\end{figure*}

QAMM/QCAM methods can also be used under such conditions, where a large number $\mathcal{O}(\sim 6 M)$ of hits are assigned to track candidates. To handle such a large number of hits, and therefore the size and number of patterns stored in the pattern library, multiple template banks are used. Using multiple pattern libraries, we can divide the detector parameter space into a finite number of cells so that the length and number of stored patterns constitute a problem suitable for the QPU device size.

To split the detector into sections where tracks with similar parameters would fall, we use a Hough Transform (HT) \cite{Duda:1972ymn}. The HT started as a technique to detect lines in an image but has been generalized and extended to detect curves in 2D and 3D. The method transforms a 2D set of points (or edges) into a hyper-parameter space where every point constitutes a line. The intersection of several lines in this new space indicates the curve's parameters connecting the real space points. The original HT utilized Cartesian coordinates to transform a set of pairs $\{(x_i,y_i)\}$ into a parameter space defined by a slope and intercept.
However, to avoid undefined slopes, it is commonly more favorable to utilize 
a polar coordinate system. In this space, the HT is built using 
\begin{equation}
\rho = x\cos\phi + y \sin\phi
\label{eq:htsp}
\end{equation}
where $\rho$ is the distance from the origin to the line. In image detection, this can be translated to twice the diagonal length of the image. $\phi$ is the angle from origin to the line. The HT method is intrinsically robust to noise, making it the perfect match for track reconstruction in noisy/inefficient detector conditions. For example, if there are noisy cells that do not correspond to the track we are attempting to reconstruct, these points will constitute a different hypothetical line in HT space, filling it with sparse, random small values, whereas the lines corresponding to the hits associated with the track will still generate a peak. Furthermore, recent studies show that this method can be implemented in hardware \cite{Abba:2014wma,Cenci:2016nrj}, allowing for faster reconstruction and as a pre-processing step for the quantum formulation for pattern matching.

A Geant4-simulated toy model for a 360 cell detector is used to study the suitability and performance of the HT method for track reconstruction. A 0.2 T magnetic field is applied in the plane perpendicular to the 2-D detector plane. To study the robustness of the method under different detector conditions, we compare the $\phi$ angle values reconstructed for several added noise values $\gamma=[0, 2,4,6,8,10,12,14,16]$ (in percentage) and simulated detector efficiencies $\eta=[95,96,97,98,99,100]$. Fig.~\ref{fig:httrack} shows the Cartesian coordinates of the activated detector cells corresponding to the track generated by a 0.5 GeV electron traversing the detector (top left). The coordinates of these eleven hits are represented as lines in the Hough space (top right) and intersecting at $\rho$ = -0.51 and angle $\phi$ = -50.0. The chosen binning for the HT space is 10 for $\phi$ and one unit for $\rho$. The value obtained for both $\phi$ and $\rho$ did not change when dropping the detector efficiency from 100 to 95$\%$, but below these conditions, the predicted angle moves from -40 to -50 degrees, as we can see in the bottom right plot in Figure \ref{fig:httrack}. In terms of performance, when noise is present, the other activated cells generate significantly more lines in HT space and therefore shift the value of the reconstructed track angle quickly. We can see this effect on the bottom left plot in Figure \ref{fig:httrack}, where the peak in $\phi - \rho$ has moved to 70 degrees, for a 15$\%$ of added noise. In this configuration, it is anticipated that the pattern bank size will contain at most six templates to match the track candidate.

Finally, in terms of future work, the HT method can be cast as a minimization problem (and be solved via quantum annealing), since the reconstructed parameters are obtained by finding the peak where most of the lines intersect.

\section{Conclusion}
This work explored using QA, run on the D-Wave 2000Q-6 processor, to perform QAMM recall and QCAM recall for track recognition.  The performance of QAMM and QCAM recall was characterized for a flexible simulated dataset, and the dependencies of the performance on pattern size, pattern density, and detector noise were determined.  To discriminate between signal and background patterns, two classification methods were used: the first based on the energy of the solution states, and the second was based on the value of a key bit designated as a classification label. We found that the performance of both QAMM and QCAM recall is strongly dependent on the pattern density and degrades for large $\alpha_s$. For small $\alpha_s$, we found that energy-based QAMM performed the best and achieved very high classification accuracy for all pattern sizes explored.  For larger track density, such as $\alpha_s = 1$ , we found that key-based QCAM performed the best.  The increase in performance for key-based QCAM as a function of pattern density is unexpected, and future work to reproduce and understand the phenomena is needed. The performance dependency on the pattern size was less pronounced and varied between pattern density, and recall and classification methods. The performance of energy-based QAMM and key-based QCAM as a function of detector inefficiency and noise was also explored. We observe robustness to detector inefficiency, but strong dependence on detector noise for all models when examining recall performance. 

As detector resolution, track multiplicity, and data rates increase in sub-atomic physics experiments, track recognition and pattern matching are becoming more computationally challenging, requiring more efficient algorithms and techniques to be explored and developed. The exploration of using quantum computing to address these challenges is relatively new, but holds promise. We found that QCAM and QAMM methods can be used to reconstruct and identify potential track candidates in collider experiments efficiently, adapting to evolving detector conditions such as malfunctioning and noise. The pattern matching method can also be extended to the clustering of calorimeter towers, and potentially to match clusters of energy in the calorimeter cells and track candidates. Future work is needed to further explore how quantum computers can be used to address these challenges.

\section*{Acknowledgements}
This work was supported by the Department of Energy High-Energy Physics program office under project 3ERKAP61.
This research used resources of the Oak Ridge Leadership Computing Facility, which is a DOE Office of Science User Facility supported under Contract DE-AC05-00OR22725. This manuscript has been authored in part by UT-Battelle, LLC under Contract No. DE-AC05-00OR22725 with the U.S. Department of Energy. The United States Government retains and the publisher, by accepting the article for publication, acknowledges that the United States Government retains a non-exclusive, paid-up, irrevocable, world-wide license to publish or reproduce the published form of this manuscript, or allow others to do so, for United States Government purposes. The Department of Energy will provide public access to these results of federally sponsored research in accordance with the DOE Public Access Plan.

\bibliographystyle{unsrt}


\end{document}